\documentstyle[preprint,revtex]{aps}
\begin{document}
\draft
\baselineskip = 1.1\baselineskip
\begin{title}
{Universal sextic effective interaction at criticality} 
\end{title}
\author{A.~I.~Sokolov, E.~V.~Orlov, V.~A.~Ul'kov}
\begin{instit}
Department of Physical Electronics, Saint Petersburg 
Electrotechnical University, \\
Professor Popov Street 5, St.Petersburg, 197376, Russia
\end{instit}
\begin{abstract}
The renormalization group approach in three dimensions is used 
to estimate the universal critical value of dimensionless sextic 
effective coupling constant for the Ising model. Four--loop 
expansion for $g_6$ is calculated and resummed resulting in 
$g_6^* = 1.596$, while the most accurate estimate for $g_6^*$ 
is argued to be equal to 1.61. 
\end{abstract}

\newpage

The critical thermodynamics of the three--dimensional Ising 
model is known to be described by Euclidean scalar field theory 
with the Hamiltonian
\begin{equation}
H = 
\int d^3 x \Bigl[ {1 \over 2} m_0^2 {\varphi}^2
 + {1 \over 2} (\nabla \varphi)^2
+ \lambda {\varphi}^4 \Bigr] \  , 
\label{eq:1}
\end{equation}
where a bare mass squared $m_0^2$ is proportional to 
$T - T_c^{(0)}$, $T_c^{(0)}$ being the phase transition
temperature in the absence of the order parameter fluctuations. 
Taking fluctuations into account results in renormalizations
of the mass $m_0 \to m$, the field $\varphi = \varphi_R \sqrt{Z}$, 
and the coupling constant $\lambda = m Z_4 Z^{-2} g_4$. Moreover,
thermal fluctuations give rise to many--point correlations 
$<\varphi(x_1) \varphi(x_2)...\varphi(x_{2k})>$ and, 
correspondingly, to higher-order terms in the expansion of the 
free energy ("effective action") in powers of magnetization $M$:
\begin{equation}
F(M) - F(0) = {1 \over 2} m^2 Z^{-1} M^2 + m Z^{-2} g_4 M^4 +
\sum_{k=3}^{ \infty} m^{3-k} Z^{-k} g_{2k} M^{2k}.
\label{eq:2}
\end{equation}
In the critical region, where fluctuations are so strong that 
they completely screen out the initial (bare) interaction, the 
behaviour of the system becomes universal and dimensionless 
effective couplings $g_{2k}$ approach their asymptotic limits 
$g_{2k}^*$, i.e. they assume constant values which are also 
universal.

Recently, several attempts have been made to find numerical values 
of higher--order coupling constants $g_6$, $g_8$, etc. at 
criticality; particular attention has been paid to the sextic 
effective interaction $g_6$. C.~Bagnuls and C.~Bervillier solving 
exact renormalization group equations in the local potential 
approximation have obtained $g_6^* = 2.40$ \cite{1}.
M.~M.~Tsypin has performed extensive Monte Carlo simulations
of the Ising model behaviour in an external magnetic field
and found $g_6^* = 2.05 \pm 0.15$ \cite{2}. T.~Reisz has deduced
the estimate $g_6^* = 1.92 \pm 0.24$ analyzing linked cluster
expansions for $O(n)$--symmetric lattice model in the Ising
case $n = 1$ \cite{3}. 

These estimates obtained by means of essentially different methods
are seen to be considerably scattered. Hence, the question about
the numerical value of $g_6^*$ remains, in fact, open. On the other
hand, there is a theoretical tool which has proven to be very
efficient when used to study the critical behaviour of various
phase transition models. We mean the renormalization group (RG)
approach in three dimensions. This method has enabled one to 
calculate critical exponents and fixed point locations with an
accuracy of order of 1 per cent for simple $O(n)$--symmetric
models \cite{4,5,6} as well as for more complicated systems
possessing two \cite{7,8,9} and three \cite{10,11} quartic
coupling constants in their Landau--Wilson Hamiltonians. It is
quite reasonable therefore to employ the field--theoretical
RG approach in three dimensions for calculation of $g_6^*$ and
other universal higher--order couplings.

In this Letter, we estimate $g_6^*$ calculating RG series for 
$g_6$ and applying Pade--Borel resummation technique to
get proper numerical results. Actually, the first step in this 
direction has been already done by one of the authors \cite{12} 
who has found two--loop RG expansion for $g_6$ which yielded
$g_6^* = 1.50$. This estimate, however, is undoubtedly very crude. 
Indeed, it has been obtained on the base of the lower--order 
perturbative expansion within the theory having no small parameter 
while accurate enough numerical estimates, as is well known, 
may be extracted only from sufficiently long RG series.
Below we find the RG expansion for $g_6$ in the four--loop 
approximation which will be shown to provide fair numerical 
estimate for the quantity of interest.

The method of calculating $g_6$ we use here is straightforward.
Since in three dimensions higher--order bare couplings are  
irrelevant in RG sense (see, e.g., Ref.~\cite{13}), renormalized
perturbative series for $g_6$ can be obtained from conventional
Feynman graph expansion of this quantity in terms of the only 
bare coupling constant -- $\lambda$. In its turn, $\lambda$ 
may be expressed perturbatively as a function of renormalized 
dimensionless quartic coupling constant $g_4$. Substituting 
corresponding power series for $\lambda$ into original expansion 
we can obtain the RG series for $g_6$. First two terms of this 
series have been presented earlier \cite{12}. Thus, what we 
should find is the three--loop and four--loop contributions. 
As can be shown, they are formed by 16 and 94 one--particle 
irreducible Feynman graphs, respectively. Their calculation gives:
\begin{equation}
g_6 = {9 \over \pi}{{\Bigl({\lambda Z^2 \over m} \Bigr)}^3}{\Bigl
[ 1 - {33 \over 2 \pi}{\lambda Z^2 \over m} + 
20.53966666 {\Bigl({\lambda Z^2 \over m} \Bigr)^2} -
73.41441479 {\Bigl({\lambda Z^2 \over m} \Bigr)^3} \Bigr]} ,
\label{eq:3} 
\end{equation}
The perturbative expansion for $\lambda$ emerges directly from the 
normalizing condition $\lambda = m Z_4 Z^{-2} g_4$ and the 
well--known expansion for $Z_4$:
\begin{equation}
Z_4 = 1 + {9 \over {2 \pi}}g_4 + {63 \over {4 \pi^2}}g_4^2 + 
1.778667825 g_4^3 + O(g_4^4).
\label{eq:4}
\end{equation} 
Combining these expressions we obtain
\begin{equation}
g_6 = {9 \over \pi}{g_4^3}{\Bigl( 1 - {3 \over \pi}{g_4} + 
1.389962951 {g_4^2} - 2.50173240 {g_4^3} \Bigr)}.
\label{eq:5}
\end{equation}

Being a field--theoretical perturbative expansion, this series 
is divergent (asymptotic). Hence, direct substitution of the fixed 
point value $g_4^*$ into (\ref{eq:5}) would not lead to 
satisfactory results. To get reasonable numerical estimate for 
$g_6^*$ some procedure making this expansion convergent 
should be applied. Since the series (\ref{eq:5}) is alternating
simple Pade--Borel technique may play a role of such a procedure.
Pade approximants of [L/1] type, when used for analytical 
continuation of Borel transforms, are known to provide rather 
good results for various Landau-Wilson models (see, e.g., 
Refs. \cite{4,6,10,11}). With the expansion (\ref{eq:5}) in hand,
we can construct Pade approximant [2/1] for its Borel transform 
$F(y)$ which is related to the function to be found 
("sum of series") by the formula
\begin{equation}
f(x) = \sum_{k = 0}^{\infty} c_k x^k = \int\limits_0^{\infty}
e^{-t} F(xt) dt , \qquad \qquad
F(y) = \sum_{k = 0}^{\infty} {c_k \over {k!}} y^k .
\label{eq:6}
\end{equation}
Numerical estimate for $g_6^*$ is then obtained by evaluation 
of the Borel integral under $g_4 = g_4^*$ where high--precision 
fixed point value of $g_4$ is known from six--loop RG 
calculations: $g_4^* = 0.988$ \cite{4,5}. The final result is 
as follows:
\begin{equation}
g_6^* = 1.596 . \ \
\label{eq:7}
\end{equation}

How close to the exact value of $g_6^*$ may this number be?  
To clear up this point let's compare the estimates given by four 
subsequent RG approximations. One--, two--, three--, and 
four--loop calculations of $g_6^*$ give 2.763, 1.500, 1.622, and 
1.596, respectively (the three--loop estimate has been presented 
earlier in Ref.~\cite{14}). Since this set of numbers demonstrates
oscillatory (and rapid!) convergence one may expect that exact 
sextic effective coupling constant lies between the three--loop 
and four--loop RG estimates. So, our four--loop RG analysis 
leading to the number (\ref{eq:7}) underestimates $g_6^*$ by less 
than 2\%. Moreover, it is possible to further improve the estimate 
for $g_6^*$ by addressing the average 
\begin{equation}
{{1.622 + 1.596} \over 2} = 1.609 \approx 1.61, \ \
\label{eq:8}
\end{equation}
which can be referred to as a most accurate approximate value 
differing from the exact one by no more than 1\%.       

Another source of information about the accuracy of numerical 
results is their sensitivity to the type of resummation procedure. 
To get such an information we calculate $g_6^*$ using the 
Borel--Leroy transformation 
\begin{equation}
f(x) = \int\limits_0^{\infty} t^b e^{-t} F(xt) dt , \ \
\label{eq:9}
\end{equation}
which contains a free parameter $b$. This parameter is chosen in 
a way that provides fastest possible convergence of the iteration
scheme. Being applied to the RG expansion (\ref{eq:5}) the 
machinery described gives $g_6^* = 1.61$ under the optimal value 
of $b$. This number is seen to coincide with the estimate 
(\ref{eq:8}) what may be considered as a strong argument in favour 
of its high accuracy. 

The value of $g_6^*$ we have just determined turns out to differ 
substantially from those obtained in Refs. \cite{1,2,3}. At the 
same time, it agrees fairly well with the estimate which follows 
from the three--loop $\epsilon$--expansion for the ratio 
$g_6/g_4^2$ \cite{15}:
\begin{equation}
{g_6 \over {g_4^2}} = 2 \epsilon - {20 \over {27}} {\epsilon}^2 + 
1.2759 {\epsilon}^3 + O({\epsilon}^4). 
\label{eq:10}
\end{equation}   
Indeed, resumming this expansion by the Pade--Borel method and 
then putting $\epsilon = 1$ and $g_4 = 0.988$ we find 
$g_6^* = 1.653$, i.e. the number which is sufficiently close to 
(\ref{eq:8}). Good agreement takes place also between our results 
and very fresh estimates $g_6^* = 1.63 \pm 0.05$ and 
$g_6^* = 1.57 \pm 0.10$ deduced from strong coupling series and 
high temperature expansions for the lattice model \cite{16}. 
At last, numbers (\ref{eq:7}), (\ref{eq:8}) are in a perfect 
accord with another very recent estimate $g_6^* = 1.604$ 
\cite{17} found by the analysis of the five--loop scaling equation 
of state \cite{18} which has used more sophisticated than ours 
resummation technique (the Borel--Leroy transformation combined 
with a conformal mapping).
    
To summarize, we have calculated the four--loop RG expansion for 
dimensionless sextic effective coupling constant $g_6$ of the 
three--dimensional Ising model. Resummation of this expansion by 
the Pade--Borel method has lead at criticality to the result 
$g_6^* = 1.596$. Having analyzed the set of four subsequent RG 
estimates for $g_6^*$ the number $1.61$ has been argued to be 
the best approximate value of universal sextic effective coupling 
accurate within 1\%. Precisely the same estimate has been found 
by applying another, Pade--Borel--Leroy procedure to resum the 
four--loop RG series. Such a stability of the obtained value under 
the change of resummation technique may be considered as a serious 
argument in favour of its high accuracy. Although $g_6^* = 1.61$ 
is substantially smaller than the early estimates \cite{1,2,3} 
it turns out to be in a good agreement with that following from 
the Pade--Borel resummed $\epsilon$--expansion and with the 
results of recent advanced calculations \cite{16,17}.
 
One of the authors (A.I.S.) is grateful to J.~Zinn--Justin for 
discussions of the results presented in Dubna \cite{14,17}. 
He also thanks P.~Butera for interesting correspondence and for 
making the paper \cite{16} available prior to publication.

\end{document}